# Mixtures of quasi-two and three dimensional hybrid organic-inorganic semiconducting perovskites for single layer LED

*Anastasia Vassilakopoulou, Dionysios Papadatos, Ioannis Zakouras and Ioannis Koutselas\**

*Materials Science Department, School of Natural Sciences, University of Patras, Patras 26504, Greece.*

\*ikouts@upatras.gr

**ABSTRACT**

New blends of simply synthesized quasi two-dimensional (quasi-2D) hydrophobic perovskite semiconductors, employed in high performance light emitting diodes (LEDs) which function due to excitonic energy transfer effects, are reported. These materials are self-assembled blends of 2D, quasi-2D and three-dimensional (3D) hybrid organic-inorganic semiconductors (HOIS). Moreover, shown for the first time, crude mixing of 3D perovskite and unprotonated amines provides similar semiconductors. HOIS reported here are based on the organic cations $CH_3NH_3^+$, $CH_3(CH_2)_7CH=CH(CH_2)_8NH_3^+$ or $C_6H_5CH_2CH_2NH_3^+$ and inorganic networks formed out of $PbX_4^{2-}$ anions (X=I, Br, Cl). HOIS exhibit strong bound excitonic states with increased oscillator strength at room temperature, tunable via simple halide substitution. HOIS blends manifest energy transfer effects, where adjacent nanoparticles of different band gap energies ($E_g$) transfer optical energy to those with the lowest $E_g$; the suggested light emission mechanism here. LED fabrication is attained via a single deposition of the hydrophobic mixture, reducing device complexity, cost and degradability. LED's diodic behavior is observed even under light albeit its photovoltaic and photoconductive quasi-2D and 3D components. LED devices exposed for over than four months under adverse laboratory conditions, showed stable light emission. Further research on this class of quasi-2D/3D HOIS mixtures is expected to lead to novel quantum electronic devices.

1. Introduction

The class of low dimensional (LD) hybrid organic-inorganic semiconductors (HOIS), see [1,2] and refs. therein, shows novel quantum phenomena, due to exciton's enhanced quantum and dielectric confinement [3,4]. HOIS span dimensionalities from 3D to 0D, including intermediate dimensionalities; those between 2D and 3D are designated as quasi-2D. The progress in classical LD semiconductors has given rise to some new types of devices [5-9], where this natural low cost class of LD semiconductors exhibits useful optoelectronic properties [4,10-15] comparable to the artificial class. Examples of such devices are LEDs based on 2D hybrid semiconductors [16-18] which function at room temperature, as well as lately reported, LEDs based on 3D hybrid materials [19-22]. These latter materials are related to this work as they share the same active semiconducting units. The LD HOIS exhibit excitons of high binding energy ($E_b$) and increased oscillator strength ($f_{exc}$), where the associated excitonic optical absorption (OA) peak can be tuned by varying stoichiometry, structure, or by simple synthetic route alteration. The increased $E_b$ is mainly due to the dielectric mismatch of the alternating inorganic and organic subparts of the HOIS [1,10]. The inorganic part is usually the semiconductor, composed of a metal halide unit network, while the organic part is usually composed of electrooptically inactive amines. In the past, a wide set of combinations of inorganic and organic parameters have led to HOIS with a variety of properties such as controlled absorption in the whole spectral ultraviolet-visible region [2], thin film transistor gate materials comparable to amorphous Si [23-25] or energy transfer optical phenomena, for example see [26-28] and *refs.* therein. 3D and quasi-2D lead halide HOIS have also been employed successfully for solar cells [29,30]. During the last two years, numerous research works related to perovskite LEDs have emerged, showing the immense possibilities of the perovskite materials due to their novel excitonic states. Important works like these following, have shown the recombination kinetic relation to excitons and defect states [31], production of fluorene free perovskite LEDs [32], green light-emitting diode from bromine based organic-inorganic halide perovskite achieving light emissions of 1500cd m$^{-2}$ [33] as well LEDs based on organometal halide perovskite amorphous nanoparticles with 3.8% efficiency [34]. Electronic properties of HOIS have been found to be partly dependent on their organic component [35], while quasi-2D HOIS have also been shown to have superb photovoltaic efficiencies [36]. Also, high performance amphiphile perovskite HOIS have been found [37], while high pressure studies have shown the intricate play of excitons with photoluminescence and structure [38] which can also can be used for laser cooling of the same systems either used in photovoltaics or LEDs [39].

In this work, blends of the presented quasi-2D/3D materials can serve as the active and single layer in LED device and it is suggested that energy transfer effects manifest in the light emission process. It is expected that such mixtures will simplify device's formation, allow for easier synthesis and manipulation of the compounds, provide resistance to the device from degrading chemicals and elucidate the light emission process. HOIS similar to those presented here, have been presented in [14] and exhibit intense optical energy transfer effects, while these can be formed even by milling together 3D with 2D or simply quasi-2D HOIS. Milling can produce compounds with tunable emission wavelength in the optical range, only by appropriate chemical tuning of the precursors' mix. For some golden ratios of 3D:2D, it is possible to yield crystalline quasi-2D HOIS of specific dimensionality, while for other ratios mixtures of quasi-2D/3D are formed, including those ascribed to the golden ratios. Each quasi-2D HOIS is associated with an excitonic, hydrogenic 1s type OA peak at room temperature, with a specific value of $E_b$ and $E_g$ [40]. The minimum $E_g$ that a quasi-2D HOIS can attain is the limit of a quasi-2D compound reaching the 3D limit, latter being the class of widely called lead halide perovskites (3D system).

The quasi-2D/3D HOIS mixtures exhibit strong distinct peaks composed of excitonic peaks in OA and photoluminescence (PL) spectra which correspond to the various crystalline quasi-2D HOIS [39]. In some mixtures, it is possible not to exhibit any discrete excitonic OA peak, except for a strong PL at the minimum agglomerate's $E_g$, *i.e.* 3D material [41]. This is a manifestation of energy transfer and such effects are not unique as in [41], but have been observed in other similar semiconductors such as $CsPbBr_3$, for example see references in [42-47]. In the latter works it is shown that microcrystalline $CsPbBr_3$ has stronger PL than in bulk or polycrystalline form. In the past, efforts as appearing in [18], for LEDs based on blends of quasi-2D/3D HOIS were unsuccessful, while now this has been attained; in the same work successful fabrication of 2D HOIS based prototype RT LEDs was reported, yet the exact model of excitonic emission was not determined. Quasi-2D HOIS systems have also been employed successfully for photovoltaic systems showing large open circuit voltage ($V_{open}$) and improved stability [48]. The authors here have preliminary tested the same LED devices for photovoltaic action, thereby proposing the reported mixtures as possible photovoltaic active materials.

Finally, it is very important to note that after this work submission new papers have appeared where 2D lead bromide LEDs have been also found by other groups [49] as well as LED based on quasi-2D systems supporting the energy transfer [50] and another work with, however, much different chemistry, phenomena and explanation of the effects than those presented here [51]. In any case, it is considered that the HOIS systems and mainly the mixtures of HOIS, since these were found [41], are bound to lead a new era of phenomena and devices.

## 2. Materials and Methods

Gallium–Indium eutectic (≥99.99% trace metals basis), Dodecylamine (DoD - grade 99% ), Phenethylamine (PhE) of grade (99%), Oleylamine (OL) technical grade (70%) as well as of higher grade (98%), Methylamine ($CH_3NH_2$,Meth, 40% in water), Indium tin oxide (ITO) coated glass (square, surface resistivity 15-25 Ω/sq), N,N'-Bis (3-methylphenyl)-N,N'-dimethylbenzidine (TPD-99%), Tris (8-hydroquinoline) aluminum ($Alq_3$ - 99.995%),

Hydrobromic acid (HBr - ACS reagent, 48%), Acetonitrile (AcN - CHROMASOLV® Plus ≥99.9%), Hydrochloric acid (HCl -ACS reagent, 37%), Hydriodic acid (HI-ACS reagent, ≥47.0%), Lead (II) iodide ($PbI_2$- 99.999% trace metals basis), Lead (II) bromide ($PbBr_2$ - 99.999% trace metals basis), Chloroform (≥99.5%), N,N'-Dimethylformamide (DMF - 99.8%), were obtained from Sigma Aldrich and were used without any further purification. The abbreviation (A-H) implies protonation of amine A. All quantities for acids refer to the previously mentioned respective solutions.

For the typical preparation, in the case of $(OL-H)_2PbBr_4$, $PbBr_2$ (0.5mmol) is dissolved in 5ml of AcN with the addition of 0.054ml of HBr and stirred until an optically clear solution is obtained. OL (1.2mmol) is dissolved as salt by stirring in 5ml AcN with HBr (0.054ml). These two solutions are slowly mixed under room temperature, yielding a dark orange precipitate. Also, such reaction can be performed with a starting molar ratio 1:1:1 for $OL:PbBr_2:HBr$, as well as ratios of the form 2:1:4. Results have been the same as far as single layer LED fabrication is concerned, irrespective of the mixing ratios. Excess of HBr was used to check its effect on the bromination of the C=C bond of the OL as well; such change was not detected by FTIR measurements. Decreasing the OL content in the 2D synthesis could lead to defects due to the extra lead halide content and such work is still in progress.

For $(OL-H)_2PbI_4$, a typical similar reaction, as the previous one, involves $PbI_2$ (0.5mmol) dissolved in 5ml of AcN with HI (0.085ml), while OL (1.2mmol) is dissolved by stirring in 5ml of AcN with HI (0.085ml), while the final solutions are mixed together, filtered and dried. The compounds prepared as $(Amine-H)_2PbX_4$ where (X=Cl, I, Br) will also be referred to as 2D semiconductors.

For the $(Meth-H)PbBr_3$, $PbBr_2$ (5.4mmol) is dissolved in 3ml of AcN with the addition of 3ml of HBr and is stirred until an optically clear solution is obtained. Methylamine (13.3mmol) is dissolved by stirring in 3ml of AcN with HBr (0.910g). These two solutions are slowly mixed and dried.

For (Meth-H)PbI$_3$, a typical similar reaction, as the previous one, involves PbI$_2$ (2.2mmol) dissolved in 3ml of AcN with HI (0.708g), while Methylamine (0.720ml) is dissolved by stirring in 3ml of AcN with HI (0.265g) and the solutions were again mixed. As these solutions are dried a black precipitate is being formed. The materials with (Meth-H)PbX$_3$ (X=Cl, I, Br) will be referred to as 3D semiconductors.

All precipitates of the previous precursors can be re-dissolved in chloroform or DMF for spin-coating or for creating mixtures. The chloride analogues are better prepared in DMF/DMSO mixtures.

The previously described compounds have also been used for creating mixtures of quasi-2D semiconductors of the formula (CH$_3$NH$_3$)$_{n-1}$(OL-H)$_2$Pb$_n$X$_{3n+1}$. It is crucial to emphasize that the previously referenced methods of mechanical milling, such as in [14], although producing strong and visible energy transfer optical effects, are not sufficient to create well connected composites, as far as current flow is concerned, in order to function the LED. For this reason, in a typical preparation the mixed semiconductors where prepared from a 4:1 weight ratio of the 3D:2D, precursors. In the case of Meth-Oley-PbBr$_4$ a typical procedure would be to mix 632mg (Meth-H)PbBr$_3$ with 158mg (OL-H)$_2$PbBr$_4$ (i.e. 4:1 weight ratio). Later, the mixture is dissolved in 2ml of DMF (sample DP1). In the case of Meth-Oley-PbI$_4$, 0.207g (Meth-H)PbI$_3$ was mixed with 66.5mg (OL-H)$_2$PbI$_4$, and dissolved in 9ml in chloroform. Similar stoichiometric protocols have been varied from a 4:1 weight ratio up to 8:1. For all reported 2D and 3D HOIS, variants with iodine and chlorine have been synthesized and finally all mixtures of 2D and 3D compounds with varying halogen content have been synthesized in the reported weight ratios.

LED prototypes can be created in a three layer fashion. ITO glass was first coated with 200μl of TPD (155mg) in chloroform (1ml) solution and spin coated at 4500rpm for 40secs. Devices were also made with a modification of the previous step at 4000 rpm for 20 secs with acceleration and deceleration, each at 20 secs. Afterwards, without heating the first layer, 200μl of the composite Meth-OL-PbBr$_4$ DMF solution was spin coated at 4500rpm for 40 secs and for the third layer, 200μl of a chloroform (3ml) solution with Alq$_3$ (0.137g) was spin coated at 4500 rpm for 40 secs. The ITO or FTO coated substrates were immersed in piranha solution for 15 minutes and rinsed with 18MΩ water. Devices exhibiting only an active layer were also fabricated. Other type of samples were created in a similar fashion where the second layer consisted of 200μl of Meth-OL-PbI$_4$ or 200μl of (OL)$_2$PbI$_4$ at a dilution of *c.a.* 5.5mg dissolved in 2ml chloroform. LED turn on was performed by manually contacting a negative power supply wire tipped with Ga-In alloy, while in other cases gold was deposited through a mask at *c.a.* 60nm thickness. Videos of the LED's emission are provided as supplementary files. Finally, interesting results have been obtained by mixing 3D perovskites with long amines, protonated or not, which yield in DMF similar results as mixing 3D and 2D HOIS. For example, mixing 0.5mmol of 3D (Meth-H)PbBr$_3$ in 600μl DMF with either 0.1mmol of OL or 0.7mg of PhE.HCl, or 0.7mg DoD.HBr yields solutions which provide composites of quasi-2D HOIS. The starting mixed HOIS solutions and LED fabrication data, all based on lead bromides dissolved in DMF unless else noted, presented in this work can be summarized in Table 1, where some data on the LED fabrication are included as well.

Heating the primary mixed solutions is important, set at about 50$^o$C, in order to create a well dissolved mixture of quasi-2D/3D compounds, which will be properly solidified upon drying the right blend as evidenced by the multitude of excitonic peaks appearing in the OA spectra. Mixing 2D systems based on amines other than OL, produced similar materials and LED devices, but the usage of OL appears to provide resistance to humidity, as well as to acetone rinsing in some cases, probably due to the outer oleic chains covering semiconducting nanoparticles. All LEDs have retained light emission for months being stored under air and light in unheated and humid laboratories.

X-ray powder diffraction data were obtained from polycrystalline samples on a Bruker D8 Advance diffractometer equipped with a LynxEye® detector and Ni filtered CuKa radiation. The scanning area covered the 2θ interval 2-80$^o$, with a scanning angle step size of 0.015$^o$ and a time step of 0.161 secs. OA spectra in the ultraviolet-visible spectral region were recorded on a Shimadzu 1650 spectrophotometer in the range of 200-800 nm, at a sampling step of 0.5nm at 1.5nm slits, using a combination of halogen and $D_2$ lamps as sources and were measured as thin or thick spin coated films on quartz substrates or on ITO substrates, after having subtracted the substrate's spectra as reference. The PL and photoluminescence excitation (PLE) spectra were obtained from solid pressed pellets or thick deposits on quartz plates or thin films, mounted in a Hitachi F-2500 FL spectrophotometer employing a xenon 150W lamp and a R928 photomultiplier. The excitation and detection slits were set at 2.5nm. For signals inducing saturation, lower accelerating voltage was chosen at 450V, from 700V. In some cases, the PLE spectra have been corrected through the instrument supplied files, estimated from materials with known quantum yields. All OA, PL and PLE spectra were recorded at room temperature, unless else noted.

Electroluminescence spectra (EL) were recorded from a proof of concept device installed at the focal length of the detection slits of the previously mentioned Hitachi fluorescence instrument, while videos of these devices functioning have been set in the supplementary information section. Recorded video and images were easily visible by naked eye and recorded for forward bias only (ITO:+), of about 2.5-7 Volts, as reverse bias did not produce any light. The EL devices have been created with and without the $Alq_3$ and/or TPD, electron and hole injection, respectively, layers. Some I-V characteristic curves were recorded using a Keithley 6517A measuring current at specified voltages with a sub-millimeter radius of Cu wire in the form of a spring slightly touching the surface of the active material (only TPD layer). The small value of currents recorded in the I-V compared to those on by manual Ga-In contact, is probably due the area of the Ga/In liquid tip. Preliminary EL quantum efficiency measurements have yielded a minimum of 3% efficiency at 530nm, however more detailed measurements are needed.

SEM/EDX measurements were performed on an EVO-MA 10 Carl Zeiss instrument equipped with a 129eV resolution INCAx-act Silicon Drift Detector. EDX were taken at 15kV and images at 5kV accelerating voltage.

## 3. Results and Discussion

### 3.1 Morphological properties of quasi-2D and 3D HOIS

The self-assembled quasi-2D HOIS in the reported synthesis are well studied in the past, while their crystalline forms have a general chemical formula $(Z)_{n-1}(M-H)_2Pb_nX_{3n+1}$ where **n**=1,2,3..; and M is an amine [52-54]. Here M can be OL, DoD, or PhE, while M-H stands for the protonated amine and Z is $CH_3NH_3$ (Methylamine-H, *abbr*. Meth-H). Typical structural schemes of these layered materials are shown in Fig. 1. When **n**=∞ (Fig. 1, right) the material is equivalent to the 3D material (aka perovskite), while when **n**=1 the material is strictly 2D. In this case, each one of the active layers, is being an infinite 2D sheet of edge sharing $PbI_6$ octahedra and thickness of *c.a.*6Å, separated by organic cations; thickness is twice the length of the Pb-halogen bond. As the inorganic network's dimensionality increases from 2D to 3D, the 2D inorganic layer's thickness increases to *c.a.* $n_x$6Å, incorporating Meth-H cations supporting its stability.

As the quasi-2D HOIS dimensionality increases, presented in Fig. 1 from left to right, the corresponding $E_g$ decreases until that of the 3D HOIS is reached. On the other hand, the behavior of excitonic $E_b$ depends on the organic molecule and the actual crystal structure. While it is hard to foresee the $E_b$ behavior with increasing dimensionality, in general, $E_b$ tends to decrease as well [10]. In most cases it is difficult to obtain crystals of a specific quasi-2D HOIS, especially for **n**>3, and thus, in such systems the OA and PL spectra appear as being composed as sum of the corresponding spectra of the individual quasi-2D/3D HOIS, see ref. [40]. For this reason, the usage of optical spectra simplifies the determination of the structural components forming in the mixture rather than relying in intricate XRD measurements and analyses.

Elemental analysis (EDX) for some of the HOIS and mixtures yielded the following molar ratios of Br:Pb under electron beam of 15kV, (Meth-H)$PbBr_3$:3.16, (PhE-H)$_2PbBr_4$: 4.35, (DoD-H)$_2PbBr_4$:4.39, (OL-H)$_2PbBr_4$:4.53, mix of (Meth-H)$PbBr_3$ and (OL-H)$_2PbPbr_4$ (4.7:1 weight ratio dried after mixed in DMF):3.5. All measurements were almost similar throughout the sample's surface, yet the values reported are averaged over three different regions. The heating and rapid cooling from DMF solution leads to uniform EDX signal from the sample surface. Simple arithmetic manipulation computes a molar ratio Br/Pb, for the previous mixed system with Meth-H and OL-H, to be 3.08. Correcting with standards for EDX, the previous ratio is estimated as 3.2±0.3, close to the measured value. Thus, the final inorganic components in the mixture are as predetermined from the precursor components; the organic part has not been analyzed. In all cases the precursors are completely dissolved in their solutions and phase separation has been observed only in cases where one of the primary components diverges from the reported weight ratios. EDX spectra for some of the samples are included in the supplementary information, Fig. S1.

Fig. 2 shows typical XRD patterns of (Meth-H)$PbBr_3$, a quasi-2D/3D mixture of Meth-OL-$PbBr_4$ (sample DP53, functioning as single layer LED) and that of the 2D material (OL-H)$_2PbBr_4$ (sample DP18). The peak appearing at $c.a.$ 2.28° (d=38Å), in Fig. 2d, corresponds to the reflection from the planes perpendicular to the long axis of the 2D HOIS's unit cell. This is in qualitative accordance with the length of the OL molecule, $c.a.$ 16Å to 20Å, which folds itself in pairs between the inorganic layers with the pair's amine groups accessing mirror inorganic layers. OL as free standing molecule is bent at the C=C bond, as computed by ab-initio geometry calculations not reported here, thus, its packing length may well be smaller than its linear length. In the past [18], the same HOIS compound was indexed in P121 space group, a=6.866Å, b=36.647Å, c=6.544Å, β=122.024°. It is presumed that small variations in a, b, c are due to the inherent instabilities of the 2D layer, attributed to the $NH_3$ head coupling to the inorganic layer. Moreover, it is possible that synthesis or re-crystallization with DMF slightly alters the unit cell [55,56]. Finally, it is noted that the actual geometrical packing of the OL molecule in the crystal structure of the associated 2D HOIS has not been determined yet and work on this is in progress.

The 3D HOIS does not exhibit XRD peaks at small angles, unlike the 2D perovskite, while the blend, DP53, shows either no significant peaks at small angles. The XRD pattern of the 3D system has been indexed with a cubic system (a=5.945Å). DP53 resembles the 3D HOIS except for some strong peaks as in 31.62° and 35.39° indicating the existence of other substructures as well. This is expected since for DP53 the molar ratio of 3D to 2D was set to 10:1. However, as it will be discussed later, DP53 OA spectra indicate the existence of quasi-2D phases. Since these phases are lamellar like, DP53 should have exhibited some low angle peaks. The fact that such are not observed is attributed to three reasons. First, the quasi-2D phases included in DP53 are probably of nano dimensions and, thus, have broad XRD peaks. Second, the inorganic part of the quasi-2D HOIS does resemble the 3D HOIS, especially for **n**>3, thus, the quasi-2D phase XRD patterns should resemble the 3D HOIS. Third, the samples were in the form of thin films, therefore, it is easy not to observe such in the standard θ°-2θ° geometry. Finally, it is worthwhile to mention that from semi-empirical calculations, as performed in [10], it has been calculated that a single 2D inorganic perovskite layer with size of 5x5 octahedral units, i.e. 30x30nm, has similar $E_g$ to that of the 2D infinite lattice. Thus, it is quite probable that the reported blends have nanosized quasi-2D HOIS, exhibiting the appropriate quasi-2D OA peaks, that are not detected by XRD.

### 3.2 Optical properties of quasi-2D and 3D HOIS and related LED properties

Fig. 3 shows the OA and PL spectra for the DP1 sample. In Fig. 3a, it can be observed that this mixture of quasi-2D/3D HOIS has specific strong excitonic OA peaks at 428nm (**n**=2 layer), 448nm (**n**=3 layer), etc, while the absorbance of the **n**>3 systems cannot be observed, except for a shoulder at 470nm (**n**=4 layer). However, the PL spectrum in Fig. 3b exhibits distinguishable peaks at 454 and 475 nm, which correspond to the quasi-2D HOIS, **n**=3 and **n**=4, respectively, and have small Stokes shifts with respect to the corresponding OA spectra. Moreover, the PL spectra in Fig. 3 shows that even though there is no relative strong visible absorbance at *c.a.* 530nm, a strong PL peak is observed at 536nm due to energy transfer effects [27]. Thus, the quasi-2D/3D mixture DP1 is comprised from a set of crystalline quasi-2D HOIS structures as well as some more undetermined structures with energy band gaps in between the possible various crystalline quasi-2D. In the top inset of Fig. 3, the still image of an electroluminescence (EL) device employing DP1 can be observed. The bottom inset image in Fig. 3 is the LED device schematic. The EL, has a peak at *c.a.* 525nm, corresponding to the 3D perovskite $E_g$. Emission of EL at higher energies was not observed nor any spreading of the EL peak due to some quasi-2D moieties with high **n**.

In Fig. 4, OA and PL spectra of the sample DP23 are shown, which exhibit slightly different behavior from that of DP1. Specifically the multitude of the strong excitonic peaks in the OA spectra, seen in Fig. 4a, at 433nm, 452nm, 471nm and 487nm correspond to four **n**-layer quasi-2D HOIS that have formed simultaneously, for **n**=2,3,4,5, respectively. The peak at 522nm is due to the OA of the exciton of the 3D perovskite. These OA excitonic positions are close in value to those of DP1. Also, each of the DP23 excitonic peaks has an associated PL peak, as shown in Fig. 4b; a Stokes shift of 6nm is observed for all PL peaks. The OA and PL peaks coincide quite well with those of DP1, suggesting the various **n**-layer nanocrystals formed have almost the same structure and dielectric confinement, which is logical since the crystalline quasi-2D HOIS are well defined.

It is important to note that all reported HOIS blends have two unique features that allow the detection of these quasi-2D phases coexisting. First, the partial organic nature of the quasi-2D, inducing dielectric confinement on the excitons, forces the optical manifestation of the excitonic peak in each phase at RT. Second, the possible crystalline phases named quasi-2D have low formation energy, thus, upon mixing 3D and 2D HOIS there is automatic self-assembly towards energetically favourable quasi-2D crystalline phases. If the blends had been another type of semiconductor but a hybrid organic-inorganic one, then most probably these phases would have gone undetected.

A device was also fabricated based on DP23 which operated at 5-8V, with less than 1mA operating current. The PL spectra of the thin film device and its EL spectrum are shown in the inset of Fig. 4. The PL of the inset is not strong because the film is very thin, however, the 3D PL exciton can be detected as a weak peak. The full width at half maximum (FWHM) of the EL spectrum is about 25nm, close the value given by researchers in [19-21] and of similar width to sophisticated cavity OLEDs [57]. It is considered to be of small width because it would be expected that this radiative recombination of the 3D exciton would be a broad peak. The reason is that the 3D exciton is coupled to a variety of recombination paths, including surface defects and local environment interactions, due to the existence of the quasi-2D components. It is unknown at this time if the narrow width of the EL is related to some form of superradiance effect, which could occur if the interacting nanoparticles are probably small enough and in such proximity that the coherence length of the exciton is preserved, allowing the coherent emission of light in PL [27] and EL at the peak wavelength of the 3D material $E_g$. No other EL peak emission is observed for the DP23 based LED at higher energies corresponding to quasi-2D or 2D HOIS.

Researchers in [47,58] have demonstrated amplified spontaneous emission in similar systems of the same order of FWHM and wavelength as for example with the DP23 based LED, an argument that supports the quantum nature of the light emission at *c.a.* 530nm. According to the latter references, the absence of traps assists the energy transfer, which is the case here as evidenced by the small Stokes shifts in between the OA and PL peaks for DP23 as well as for the DP1 blend.

Although Fig. 4a OA peaks at 433nm, 452nm, 471nm seem weak in intensity, this does not imply that the number of the respective quasi-2D units is small. The reason is that the aforementioned OA peaks are appearing on top of a broad absorption profile, which is not to be taken as a baseline, and the peaks seen are due to the large percentage of the crystalline quasi-2D HOIS. It is also suggested that the film is comprised, also, of other halide units not resembling the quasi-2D with (**n**=1, 2, 3...) but of $E_g$ values close to the crystalline quasi-2D.

In Fig. 4, the PL and PLE spectra of a dense DP23 solution in DMF are provided as dotted lines. These spectra exhibit broad and few distinct peaks since it is common knowledge that a multitude of perovskite nanostructures form in DMF and other solvents upon dissolving lead halides [59]. These units, assisted here by the diluted organic components, give rise to a broad OA (comparable to PLE) and PL spectra. It is expected from the corresponding PLE peak, 397nm in Fig. 4c, that there are 2D-like nanoparticles in the solution, associated with a broad PL peaked at 417nm.

OA, PL, EL spectra for the samples DP53 and DP80 are presented in Fig. 5. The OA spectrum for DP53 in Fig. 5a exhibits strong OA excitonic peaks for the **n**=2 quasi-2D at 430nm and that of the 3D perovskite at 521nm. DP53 exhibits PL peaks, Fig. 5b, corresponding to quasi-2D moieties at 437nm, 456nm, 475nm, 491nm and only a weak PL peak at 530nm. For the DP53 based LED, 200μl of hot solution, once spin coated, provided a LED functioning with voltages from 3.5V up to 20V with stable green light being emitted, with no hole and electron injection layers (supplementary information video). Also, it is quite interesting that the DP53, exhibits a hardly detectable PL peak at 532nm in contrast to DP23 and DP1 samples, yet it does show strong EL signal. The inset left image in Fig. 5 shows the color of DP53 when excited by 404nm radiation, which is blue-green due to the combined strong PL of the quasi-2D and the 3D material. In the right inset an image of the green LED with DP53 is shown. There is no light emitted in other wavelengths than the narrow peak at 532nm for the DP53 EL spectrum.

DP80, which is a 1:1 weight ratio of 3D to 2D blend, functioned as LED with turn-on voltages of 2V up to 13V (see supplementary information for video of LED). The EL emission, Fig. 5e, exhibited peak centered at 533nm. DP80 shows, in Fig. 5d, OA excitonic peaks at 401nm, 429nm, 452nm and 521nm, which correspond to the **n**=1, **n**=2, **n**=3 and **n**=∞ quasi-2D HOIS, respectively. The first of these peaks, at 401nm, is associated with the 2D HOIS, which is well understood to appear since the 3D:2D mole ratio material is almost two;

thus, there is enough organic material to preferentially synthesize the 2D ($n$=1) LD HOIS. DP80 was not easily spin-coated on ITO substrate, unless TPD was used.

An amazing capability of the LD HOIS is their strong tendency to self-form once the appropriate chemical environment appears. This tendency was implemented for forming the DP106 and DP112 samples. LEDs incorporating LD HOIS DP106 and DP112, function as a single layer LED (supplementary information video for both sample). In fact, these active layer blended HOIS were synthesized by injecting a long amine salt, or a pure amine, for example OL, PhE, or DoD or their halide salts, in the DMF solution of (Meth-H)PbBr$_3$. The final solution was later spin coated on ITO, in some cases without even treating ITO with piranha solution. There is a possibility that this method is conceptually related to doping perovskites LED with CsBr as in [60]. Here for DP106, the amine, not its salt, transforms the inorganic network into a multitude of quasi-2D nanoparticles.

This effect is explicit in the OA and PL spectra of Fig. 6a and 6b for sample DP106, which is a 3D perovskite injected with pure OL. The OA spectrum, Fig. 6a, shows peaks at 431nm ($n$=2), 451nm ($n$=3), 473nm ($n$=4) and 523nm, which correspond to some of the quasi-2D HOIS OA excitonic peaks as well as the absorbance peak of the 3D perovskite. In Fig. 6b there are PL peaks associated with each OA peak, with variable Stokes shifts of a maximum 9nm. The 3D PL peak appears to be composed of a twin peak probably associated with defects. The EL spectrum of DP106, not presented here, showed a broad peak centered at both the PL peaks, however, the resolution was not sufficient for presentation. It is possible that DP106 is bromine deficient, thus, it appears that as in the related works [20-22] bromine doping is important; further research is needed on this subject.

Fig. 6c shows the OA of the blend DP112 which exhibits no distinct excitonic absorbance peaks but the one of the 3D perovskite at 518nm. DP112 exhibits, however, PL peaks as seen in Fig. 6d, of the 3D compound and three peaks corresponding to three $n$-layer quasi-2D. These PL peaks are at 407nm, 437nm, 458nm for the quasi-2D of $n$=1, $n$=3, $n$=4, respectively. The 3D PL peak for DP112 appears at 526nm. All these data demonstrate that there are quasi-2D HOIS composing DP112 which do not show up as peaks in absorption spectra, however, their existence appears in the PL spectrum. All the comparisons between PL and OA until now indicate that the PL technique is more sensitive in exposing the different quasi-2D phases in these HOIS blends. EL for DP112 is only provided as video in the supplementary data.

Fig. S2 (supporting information) displays a characteristic I-V curve measured for the DP53 based LED device. The signs of V and I in the graph have been reversed such that at *c.a.* 3V light is emitted from the device on electron flow from the Cu wire into the active material. At 1V the current measurement shifted from 0.2nA to 40nA upon light illumination. There is a strong possibility that the device's own light would affect its LED behavior. It is proposed that this phenomenon could be used to create a photonic switch where light can control the emitted light or even turn the LED on. In this context, since the quasi-2D and 2D perovskites show strong photoconductivity signal [40], it is possible that external light would be able to alter/initiate the electron pathways, thus, making a perovskite photonic switch a reality. The small observed currents are due to the submillimeter probe attached to the sample surface, thereby accessing a small surface area.

Fig. S3 (supporting information) portrays the morphology of the 3D, 2D and quasi-2D DP53 blend as seen by Scanning Electron Microscopy (SEM). The 3D perovskite appears with a standard morphology of an interconnected semiconductor, while the 2D HOIS appears as small platelets within a rubber-like substance. Normally the 2D HOIS, not based on OL, appear as smooth, clean, flat crystals, however, OL appears to alter this morphology. Finally, the mixed quasi-2D/3D semiconductor exhibits an intermediate of the 2D and 3D morphology. Similarly DP106 is observed in SEM (Fig. S4) after heating at 115$^{\circ}$C, where the intermediate morphology is again observed.

As far as tuning the wavelength of the EL peak, the iodine and chlorine analogues of the reported blends of quasi-2D/3D as well as mixed halogen stoichiometries of all I, Br, Cl analogues have been prepared. The iodine composites, however, are not stable in air but some devices were observed with naked eye to emit dark red light, but due to their fast degradation such spectra were not recorded. Ga/In may be well inducing deterioration of the iodine containing film. The pure chlorine analogues did not provide light emission to the best of our experiments; however, it has been possible to emit light at other wavelengths by appropriate mixing of the lead bromide compounds with other halogen atoms. For example, emission at 500nm is achieved by mixing phenethylamine hydrochloride (PhE.HCl) with 3D lead bromide perovskite. The EL image can be seen as inset in Fig. 7 along with the respective OA and PL spectrum. The mixing with PhE.HCl has transformed the 3D lead bromide compound towards a homologous compound with larger energy gap, which shows no absorption at 500nm, yet does exhibit PL at 500nm, through energy transfer effects. Similarly the electroluminescence is provided only at the 3D PL peak, as seen in the inset with color being blue-green. In general, it is possible that the synthetic routes to achieve the bromide quasi-2D/3D HOIS mixtures presented here may not be appropriate for all possible combinations of mixing 2D and 3D of varying halogen content, neither for mixing all possible combinations of bromide 3D perovskites with 2D or amine salts containing I or Cl. We speculate that the actual nanocrystalline units and their connectivity play an important role in the EL spectra and thus, it is explained why these materials upon heating do not continue to function well as EL devices. Also, it is probable that heating disrupts the 2D inorganic layer connectivity as the organic layer becomes disordered and expands. Finally, the connectivity of the active octahedral units as forming layers most probably play a role in the EL mechanism.

### 3.3 Electroluminescence Model Analysis

In [26] mechanical milling was shown to be a method able to produce mixed blends of quasi-2D/3D semiconductors manifesting optical energy transfer effects for all mixed halogen stoichiometries, controllably covering most of the optical spectrum. It appears that optical luminescence coupled to energy transfer phenomena is easily attainable by simple sample manipulation. On the other hand, for the same blends, mechanical milling could not lead to an observable EL signal. It appears that energy transfer is present only if the materials are dense and well connected, since real electrical paths are needed to create the excitons by electron and hole flow through the material. Such electrical connectivity could not be attained in the past by simple mechanical contact of the mixed material by positive and negative electrical contacts of different work function.

The experimental UV, PL and EL spectra presented here for blends of quasi-2D/3D HOIS show that these materials can be incorporated successfully in single layer LEDs. Regarding the model of light emission, first, it is deduced from the peaks appearing in the all composite's OA spectra, that EL at *c.a.* 530nm is possible from mixtures that exhibit or not appreciable OA component at 530nm with respect to the 2D and quasi-2D peaks; *i.e.* 3D absorption is less than the OA of the quasi-2D components in the same blend. This shows that the insertion of the 2D system induces modification on the 3D and plays a role in the EL emission at the $E_g$ (3D). The quasi-2D are real phases in the crystal as revealed by the existence of the excitonic peaks in respective OA and PL spectra. Thus, it can be ruled out immediately that, for example, the quasi-2D phases formed are just an inactive matrix supporting the 3D perovskite.

A model, named for short DRIFT, would describe the blend as being composed mainly of 3D like nanoparticles that act as electron and hole radiative recombination sinks by virtue of their small $E_g$. In DRIFT, the quasi-2D cannot be just a current sustaining matrix, or else it would have short circuited the +/- poles disabling electron flow over the 3D crystallites; this would have also appeared in increased leakage current values.

It is not possible, either, to exclude out completely the function of the 2D and quasi-2D nanoparticles as being recombination sinks; envisioning that these nanoparticles do not behave as quasi-2D nanoparticle electron and/or hole sinks, it would be logical to assume the quasi-2D crystalline nanoparticles (NPs) as adjacent to the 3D NPs; in fact this is how quasi-2D are placed since we observe PL enhancement through optical energy transfer effects.

Then, the EL of the 3D NPs occurs by carrier injection from the surrounding matrix, which could be, for example at some point, an **n**=1 2D NP. Should this **n**=1 NP be adjacent to an **n**=4 quasi-2D NP as well, it would have provided the latter with a sheer carrier injection, yielding EL from the **n**=4 quasi-2D NP. Therefore, it would be expected to have an EL signal from a quasi-2D moiety, besides the 3D perovskite, which would appear at a higher energy than that corresponding to the $\lambda_{PL}$=530nm, i.e. 2.33nm. The reason is the relation of the $E_g$ for the **n**-layer quasi-2D systems: $E_{g,\mathbf{n}=1} > E_{g,\mathbf{n}=z} > E_{g,\mathbf{n}=z+1} > E_{g,\mathbf{n}=\infty}$; assuming the absolute energy level's alignment supports that the LUMO for **n**=z is lower than the LUMO for **n**=z+1 for all quasi-2D HOIS. However, such a higher energy emission was not been observed. It is possible that the percentage of active units in the quasi-2D and 2D nanoparticles is less than in the 3D material, since the former are also composed of bulky OL cations, thus, the intensity of emitted light is too weak for observation.

An argument in favor of DRIFT could be the fact that the most carrier sinks are the 3D material, thus, radiation appears mainly at 530nm, since the 3D HOIS part would be expected to be an order of magnitude more than the 2D part. But on the other hand, the OA and PL excitonic peaks of the quasi-2D phases are strongly evident in OA spectra and this indicates that the quasi-2D species form as a substantial number of units within the mixtures. Moreover, blend DP80 is in mole ratio of 3D:2D to 2, yet still displays LED emission only at 533nm. Moreover, within DRIFT, it would be expected that in the region between the poles, there would exist many serially connected triplets of quantum wells of the type "high $E_g$-low $E_g$-high $E_g$" HOIS. This would likely hinder the motion of electrons and holes by placing them in far-apart energy minima, thus, the emission from the low energy gap HOIS would have been hindered.

It is noted that LED action occurs at particular polarity only (Ga-, ITO+). If light emission was due to DRIFT then that at high voltages (strengthening injection) any polarity would give off light, since the quasi-2D would probably provide the random media for injection and carrier drifting; such reverse emission was not evidenced either but only an electric breakdown of the film.

In the past, 2D HOIS prototype room temperature LEDs [18] were fabricated and studied, as a proof of concept, and EL was observed with naked eye for OL based 2D systems. It was possible, unpublished though, to have eye-observable emission with the system TPD/4-fluorophenethylamine lead iodide/Alq$_3$ will be reported elsewhere. Experiments in [18], where all compounds were extremely well dried, pointed to the fact that the device geometry, the coating method and material thickness were crucial to current flow and light emission. The emission of light in those LEDs, observed at high voltages, exhibited large Stokes shifts in some cases *wrt* to the exciton OA peak and were not linked to any atomic emissions possibly arising from some form of high voltage breakdown. Regardless of the emission center, there was proof that ITO and Ga/In are sufficient to induce emission from similar to those presented here 2D HOIS at room temperature. Thus, in the spectra presented in this work, EL should have been observed originating from some 2D or quasi-2D HOIS; however, no such emission was observed either. The absence of EL spectrum at higher energies than the 3D E$_g$ indicates to that some form of energy transfer possibly manifests in these materials, possibly combined with some form of superradiance.

Some blends of HOIS, such as DP1, do not have significant strong OA at the 3D E$_g$. However, they do exhibit PL and EL at the 3D E$_g$, again with no EL emission appearing at high energy levels corresponding to a 2D or a quasi-2D semiconducting species. Therefore, since the OA signal at the 3D E$_g$ is minute, it would be expected to observe EL at other higher energies than the 530nm peak, or at least observe a broadened EL signal, which was not detected either. It was not possible to detect other EL, than the 530nm peak, at all.

Following the above discussion, an extension to DRIFT is proposed, to account for the lack of emission at the quasi-2D and 2D excitonic energies, especially for the high **n** quasi-2D. The carriers, drifting in excited states, can attach to specific nano-entities (3D, quasi-2D), from where it is possible to recombine with two distinct paths: a) from different dimensionality particles (which should have displayed broadened EL spectra), or b) attach to the same nanoparticle forming an exciton [61]. In these particular quasi-2D semiconductors more novel phenomena can also be observed with many body interactions taking place, which may account for some of the phenomena seen here [62]. In the last path described before, the injected exciton could either i) recombine, ii) drive nearby excitons to recombination or iii) transfer its excited dipole energy to nearby nanoparticles, possibly with some Förster resonance energy mechanism. All three processes may not have the same probability and depend on the final composition of the mixture. Should these (ii) and (iii) processes occur in the quasi-2D materials, then these processes are bound to transfer their energy to the lower $E_g$ HOIS, which is the methylamine lead bromide perovskite. Also, the excitonic oscillator strength of the 2D and quasi-2D is quite larger than in the 3D materials due to the dielectric confinement, thus, the 2D/quasi-2D excitons probably have the capacity to affect their surrounding particles. Therefore, in the above analysis of pro and against arguments, it is suggested that the LED mechanism with the quasi-2D materials is most probably related to excitonic energy transfer.

Last, in this refined model, an explanation must be provided for the existence of the composite's PL peaks attributed to the quasi-2D HOIS. Within the model of optical energy transfer, it would be presumed that the main PL intensity would appear at the 3D material $E_g$.

This PL energy transfer is seen for the LED based on DP1, where PL is observed for the 3D entities despite the corresponding weak visible absorption observed as small peaks at the 3D exciton position, thus, energy is transferred. However, in DP1, the **n**=3 and **n**=4 quasi-2D PL peaks are still observed but no corresponding EL emission is seen. Thus, it is concluded that the PL energy transfer is not as efficient as the EL energy transfer. It could also be a possibility that the energy transfer mechanism for PL differs from that of the energy transfer for the EL and may have to do with the statistics of the positional distribution of quasi-2D with respect to that of the 3D nanoparticles.

Most important, in order to test the above theory of energy transfer as well as our methods we have experimented at 77K. As a model test we prepared a film of (4-fluorophenethylamine)$_2$PbI$_4$, which is a 2D HOIS, exhibiting excitonic absorption and PL emission at c.a. 510nm. This, as a single-layer LED active material, displays naked eye visible green electroluminescence, upon device cooling at 77K as it has been reported in [17] as well as at RT, first reported in [18]. The 2D systems show EL effortlessly at 77K. Similarly, as an analogy to the previous mentioned iodine HOIS, the blends presented here were tested at 77K. The tests did not reveal any EL color other than the 530nm emission, to the best of our knowledge and efforts. Should there be no energy transfer effects we would have expected to see EL at higher energies than the reported peaks centered from 520 to 530nm and in particular at the quasi-2D excitonic PL peaks. This should have occurred due to the multiple 2D species coexisting in the mixtures, since it has been reported that at least at 77K the 2D HOIS, presumably the quasi-2D as well, would display EL at the absorption wavelength of the excitonic peaks. It is noted that EL for the 2D type compounds that has been observed in [18] at RT was not centered at the absorption wavelength of the 2D excitons and it was neither related to any atomic emission peaks due to any moieties of the film being decomposed. Therefore, concluding, it is expected that EL reported here is related to energy transfer effects. It is interesting to check in future if the low T induces some type of grain formation on the 2D platelets allowing it to emit at the $\lambda_{exc}$. Video of 77K LED device turned on is found in the supplementary section, which includes demonstration of its PL enhancement when cooled to 77K.

Lately, a new approach to perovskite LEDs appeared [22] where PEO is being embedded with 3D HOIS (perovskite) microcrystals, providing strong EL with recombination at 550nm and Stokes shift with respect to the corresponding OA of about 20nm. Despite the similarity with the EL spectra presented here, the mechanism suggested is not the same as in [22], as for example, in blend DP1. In addition, in blends such as the DP106 and DP112, quasi-2D systems were observed to form by decomposition of the 3D perovskite by virtue of the added amine. These formed quasi-2D semiconductors are directly related to the increase of the EL of the perovskite material, which has been observed qualitatively only. It is interesting to investigate if such quasi-2D formation can occur within other organic materials or matrices other than amines or even due to some degradation of the perovskite, such as humidity.

Finally, a large set of yet unexplored crystalline hybrid organic-inorganic low dimensional semiconductor structures, for example seem [63,64] as well simple chemical synthesis tweaks [65] are envisioned as possible hybrid semiconductor candidates that may lead to the fabrication of novel quantum devices.

## 4. Conclusions

A new set of blends of nanocrystalline materials for exploring quantum phenomena in light emitting diodes is proposed, which are mixtures of quasi two and three dimensional lead halides, mainly bromides. This can also be achieved by doping the 3D perovskite with amine. These LEDs can be fabricated as a single layer device between ITO and Ga/In electrodes, even if applied manually as thick coating. In comparison to the pure 3D perovskite active materials, under the same laboratory procedures, quasi-2D/3D mixtures provided simpler and robust LED fabrication as well as increased light intensity. The LED emission profile peaked in most samples at 530nm, with FWHM of *c.a.* 30nm. It is suggested that energy transfer phenomena take place, linking to some form of superradiance effect. These effects arise from the set of NPs containing large $E_g$ HOIS (decreased dimensionality) transferring their energy towards those of smaller $E_g$ (increased dimensionality), the smaller $E_g$ being that of the 3D perovskite. It is suggested that by further tuning of the individual organic components' physicochemical properties, novel semiconductors can be obtained, for example being hydrophobic. Some research on similar novel materials for LED and related energy transfer effects will be published elsewhere.

**Acknowledgment**

We would like to thank our colleague Dr. M. Vasileiadis for the Table of Contents image.

**Electronic Supplementary Information**

Videos of the LED based on DP53, DP80, DP80 at 77K, DP106 and DP112 are in the supplementary section as MP4 files, with the sound redacted. I-V characteristics curve of DP53 is also in supplementary information along with SEM images of the 2D, quasi-2D/3D and 3D perovskite materials based on lead bromide units to elucidate their micrometer range morphology. Moreover, EDX spectra have been added for some of the semiconductors presented in the text as well as in the SEM images in the supplementary section.

**AUTHOR INFORMATION**

Corresponding Author

*E-mail: ikouts@upatras.gr

**Notes**

The authors declare no competing financial interests.

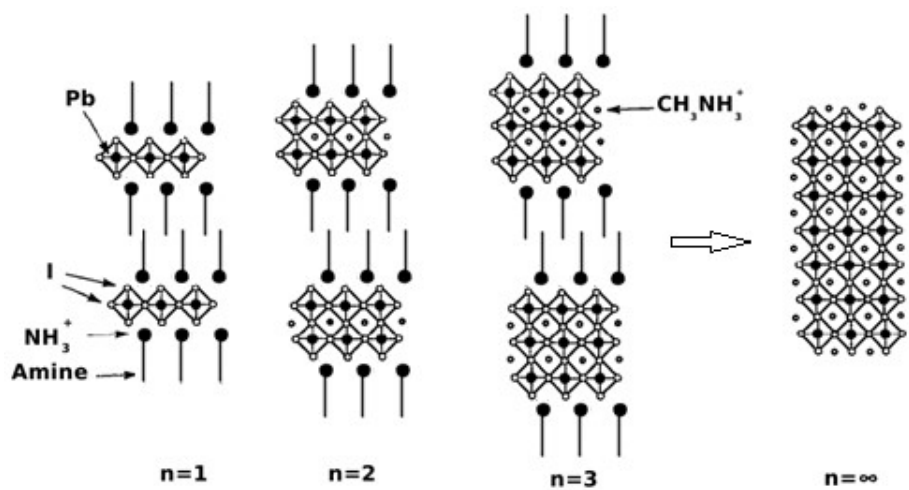

**Fig. 1.** Schematic representation of the layered structure for the crystalline $(CH_3NH_3)_{n-1}$ (OL-H)$_2$Pb$_n$X$_{3n+1}$ LD HOIS. Dimensionalities range from 2D to 3D, for n=1,2,3,.. ∞. The active layer (n=1, left) becomes progressively thicker until it reaches the 3D (n=∞, right) perovskite.

**FIGURE 2**

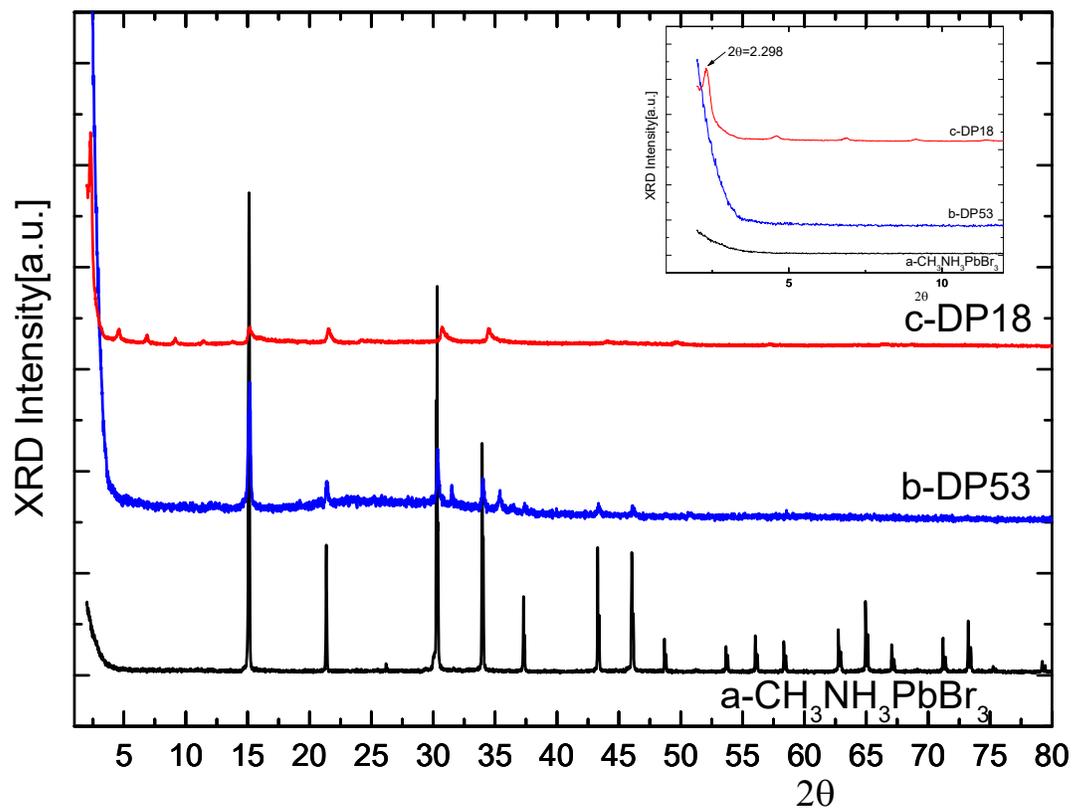

**Fig. 2**. Powder XRD patterns for the HOIS: (a) 3D (Meth-H)PbBr$_3$, (b) quasi-2D DP53, (c) 2D DP18. Inset shows the same patterns zoomed in the low angle region 1-12°.

**FIGURE 3**

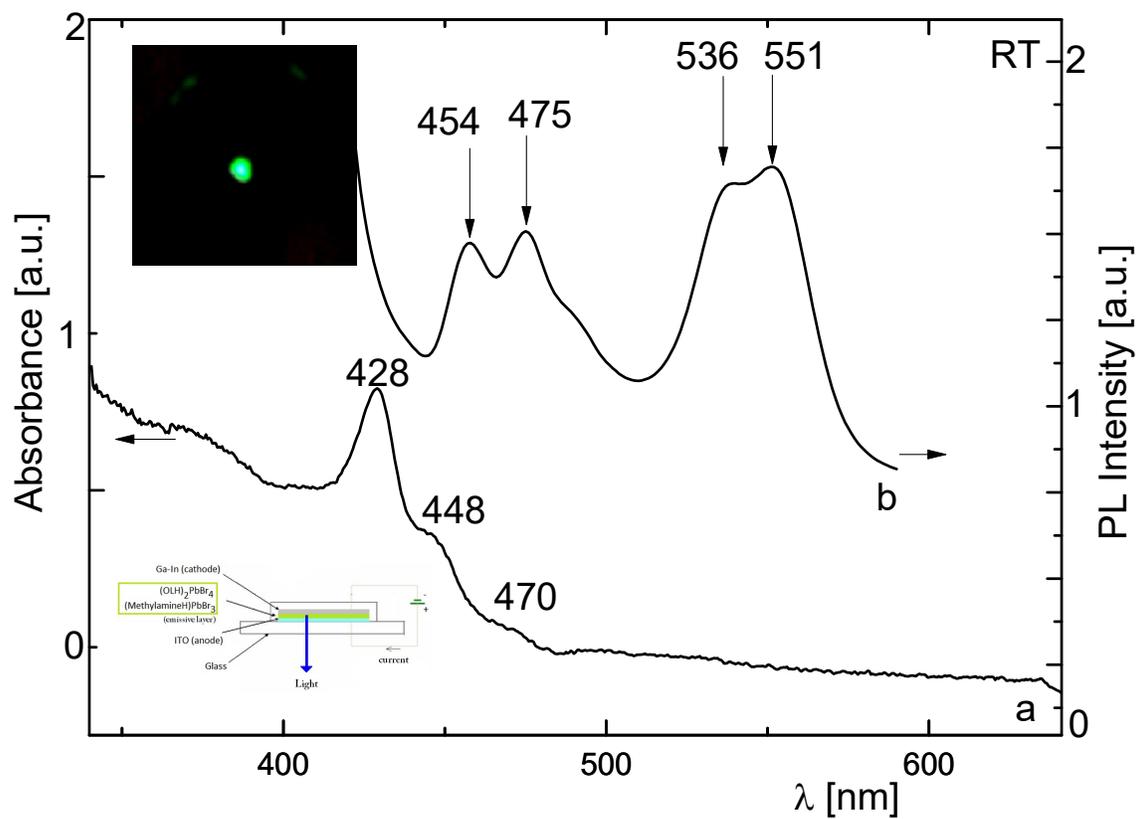

**Fig. 3**. OA (a) and PL (b, $\lambda_{exc}$=350nm) spectra for DP1 in thin deposit form. Insets, top: DP1 in LED, bottom: LED schematic.

**FIGURE 4**

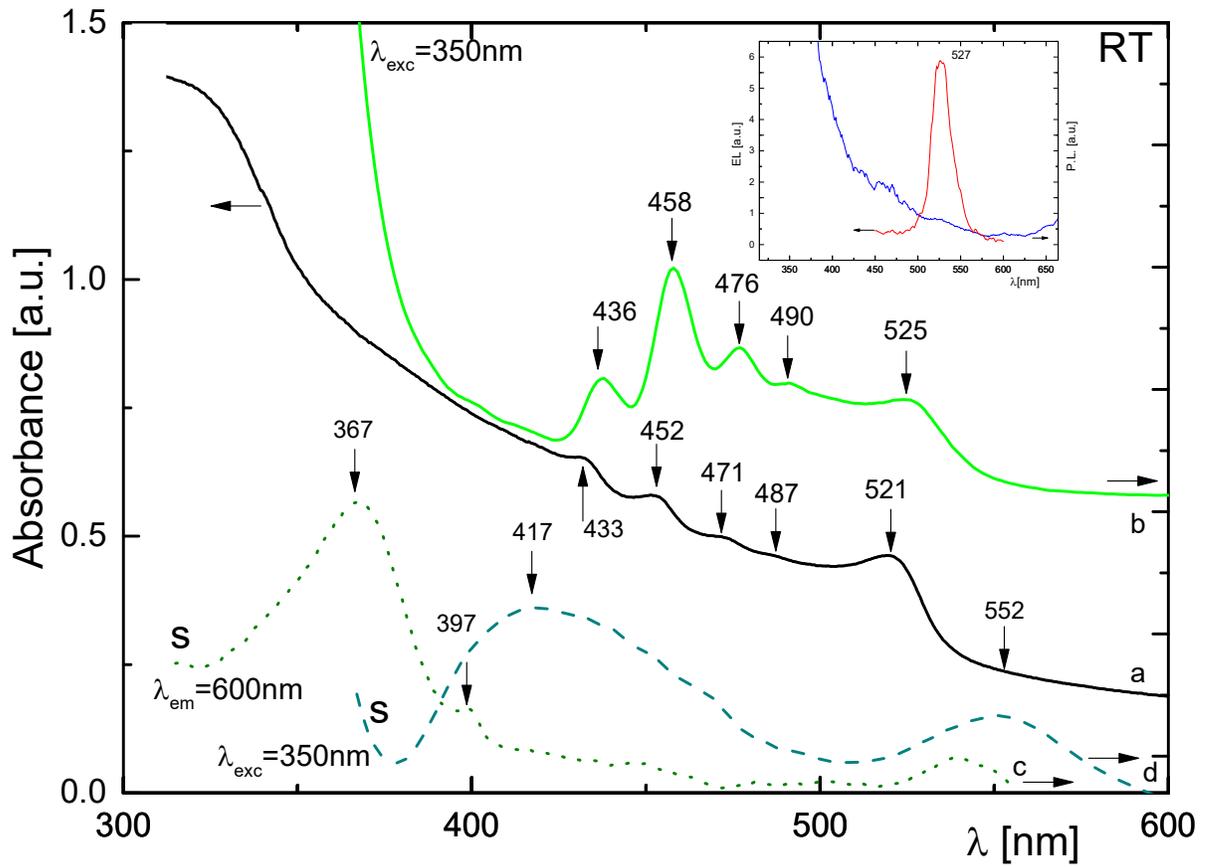

**Fig. 4.** OA (a) and PL (b, $\lambda_{exc}$=350nm) spectra for DP23 in thin deposit form; PLE (c, $\lambda_{em}$=600nm) and PL (d, $\lambda_{exc}$=350nm) spectra of the corresponding DMF solution. Inset shows the EL spectrum (left axis, red) of the LED device based on DP23 as well as its PL spectrum (right axis, blue).

FIGURE 5

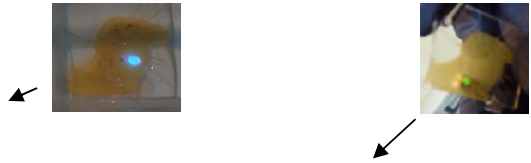

**Fig. 5**. OA (a, d), PL (b, $\lambda_{exc}$=300nm) and EL (c, e) spectra for the DP53 (a, b, c) and DP80 (d,e) in thin deposit forms. Inset images, left: DP53 illuminated with 404nm, right: single layer DP53 based LED operating at 4V.

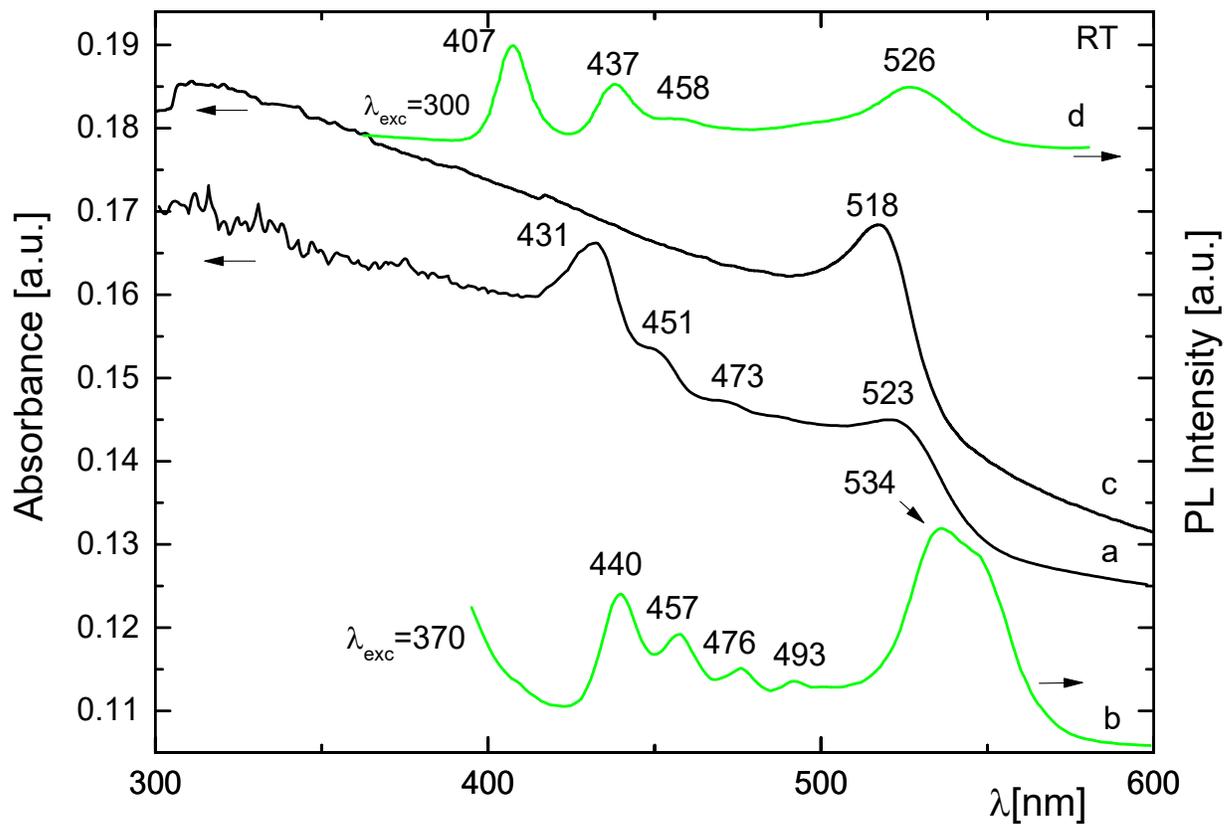

**Fig. 6**. OA (a, c) and PL (b, d) spectra for DP106 (a, b) and DP112 (c, d), where $\lambda_{exc}$ is for (b) 370nm and for (d) 300nm.

**FIGURE 7**

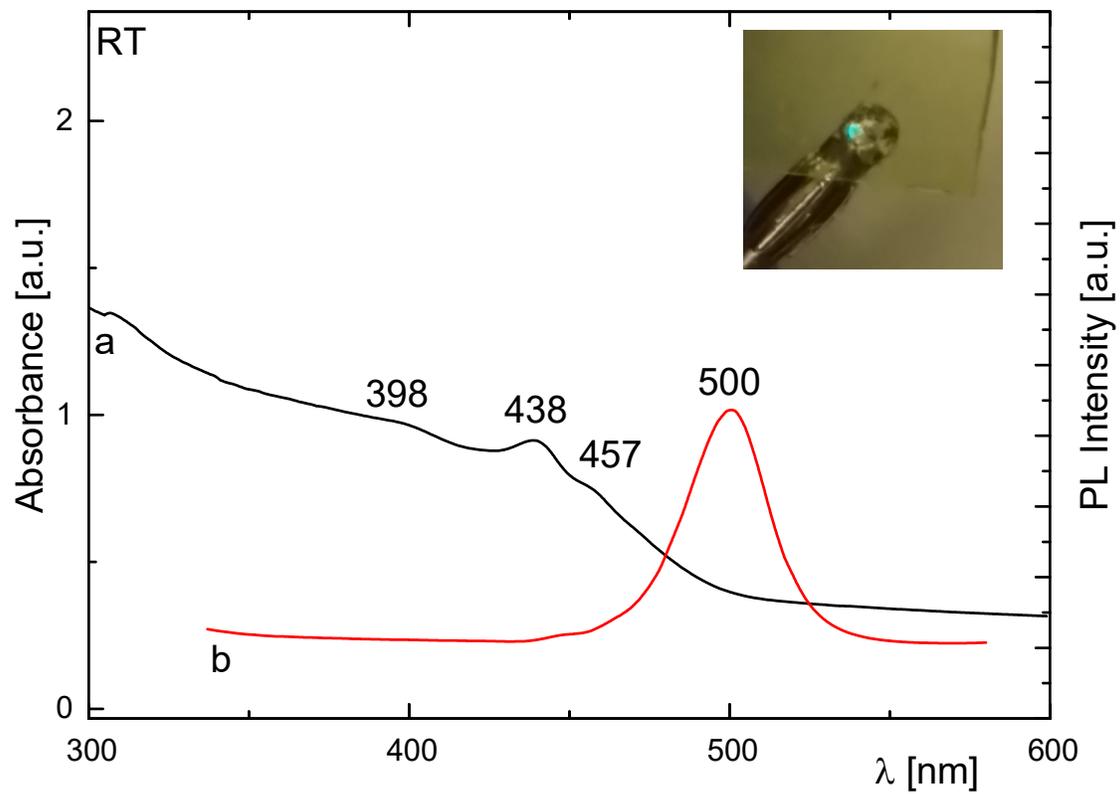

**Fig. 7.** OA (a) and PL (b) spectra for blend of PhE.HCl and 3D lead bromide perovskite in DMF with weight ratios of (7:48:200), respectively. Inset shows LED operation, emitting in light blue-green color.

**FIGURE CAPTIONS**

**Fig. 1.** Schematic representation of the layered structure for the crystalline $(CH_3NH_3)_{n-1}$ (OL-H)$_2$Pb$_n$X$_{3n+1}$ LD HOIS. Dimensionalities range from 2D to 3D, for n=1,2,3,.. ∞. The active layer (n=1, left) becomes progressively thicker until it reaches the 3D (n=∞, right) perovskite.

**Fig. 2.** Powder XRD patterns for the HOIS: (a) 3D (Meth-H)PbBr$_3$, (b) quasi-2D DP53, (c) 2D DP18. Inset shows the same patterns zoomed in the low angle region 1-12°.

**Fig. 3**. OA (a) and PL (b, $\lambda_{exc}$=350nm) spectra for DP1 in thin deposit form. Insets, top: DP1 in LED, bottom: LED schematic.

**Fig. 4.** OA (a) and PL (b, $\lambda_{exc}$=350nm) spectra for DP23 in thin deposit form; PLE (c, $\lambda_{em}$=600nm) and PL (d, $\lambda_{exc}$=350nm) spectra of the corresponding DMF solution. Inset shows the EL spectrum (left axis, red) of the LED device based on DP23 as well as its PL spectrum (right axis, blue).

**Fig. 5**. OA (a, d), PL (b, $\lambda_{exc}$=300nm) and EL (c, e) spectra for the DP53 (a, b, c) and DP80 (d,e) in thin deposit forms. Inset images, left: DP53 illuminated with 404nm, right: single layer DP53 based LED operating at 4V.

**Fig. 6.** OA (a, c) and PL (b, d) spectra for DP106 (a, b) and DP112 (c, d), where $\lambda_{exc}$ is for (b) 370nm and for (d)300nm.

**Fig. 7.** OA (a) and PL (b) spectra for blend of PhE.HCl and 3D lead bromide perovskite in DMF with weight ratios of (7:48:200), respectively. Inset shows LED operation, emitting in light blue-green color.

**Table 1**. Information on i) the stoichiometry of each active material solution (before spin coating) and ii) the respective LED structure involved. The right column refers to incorporation of electron/hole injection layers(IL). (TPD hole IL, $Alq_3$ electron IL). The following abbreviations hold 3D:$CH_3NH_3PbBr_3$, 2D:$(CH_3(CH_2)_7CH=CH(CH_2)_8NH_3)_2PbBr_4$, Meth:methylamine, OL: oleylamine, PhE: phenethylamine, DoD: Dodecylamine.

| Sample | HOIS Weight ratio 3D:2D | Solution content in 3D and 2D HOIS | DMF (ml) | Organic part analysis | e/h Inj. Layers |
|---|---|---|---|---|---|
| DP1 | 4:1 | 3D:632mg, 2D:158mg | 2 | 3D:Meth, 2D:OL | TPD,$Alq_3$ |
| DP23 | 5.6:1 | 3D:790mg, 2D:158mg | 3 | 3D:Meth, 2D:OL | TPD |
| DP53 | 4.7:1 | 3D:807mg, 2D:169mg | 1.5 | 3D:Meth, 2D:OL | *None* |
| DP80 | 1:1 | 3D:102mg, 2D:102mg | 1 | 3D:Meth, 2D:PhE | *None* |
| DP106 | *Not applic.* | 3D:239mg + 32μl OL | 0.6 | 3D:Meth, OL | TPD |
| DP112 | *Not applic.* | 3D:55mgr + 0.7mg PhE.HCl *or* 0.7mg DoD.HBr | 0.2 | 3D:Meth, PhE.HCl *or* DoD.HBr | TPD *or* *None* |

**FIGURE S1**

**EDX for 3D perovskite HOIS: (Meth-H)PbBr₃**

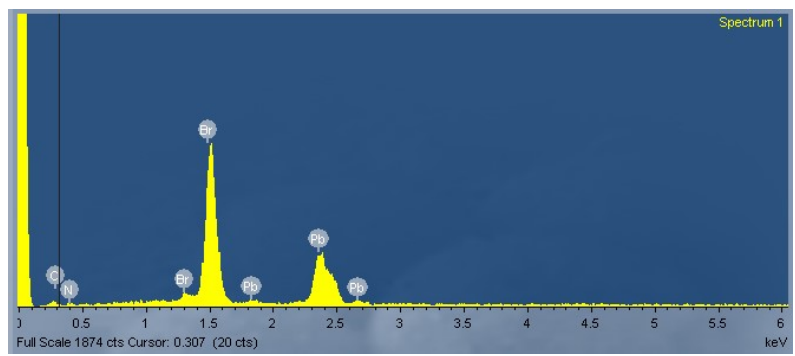

**EDX for 2D HOIS: (Phenethylamine-H)₂PbBr₄**

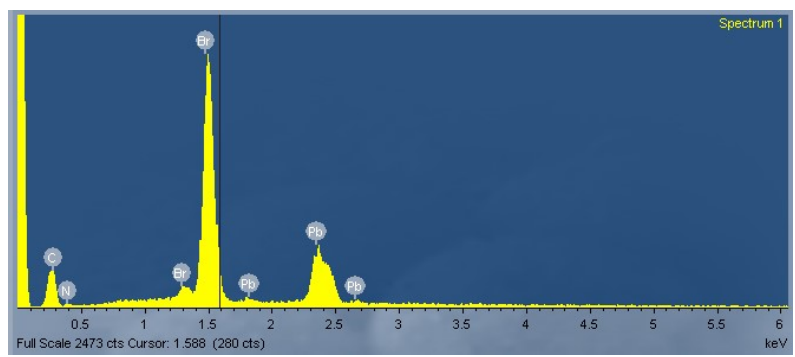

**EDX for 2D HOIS: (Oleylamine-H)₂PbBr₄**

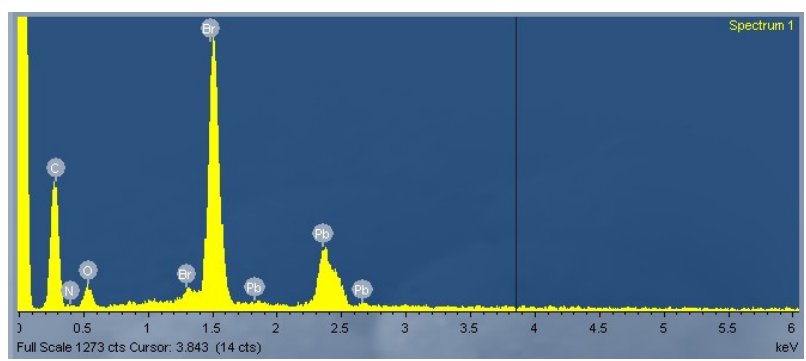

**EDX for 2D HOIS: (Dodecylamine-H)$_2$PbBr$_4$**

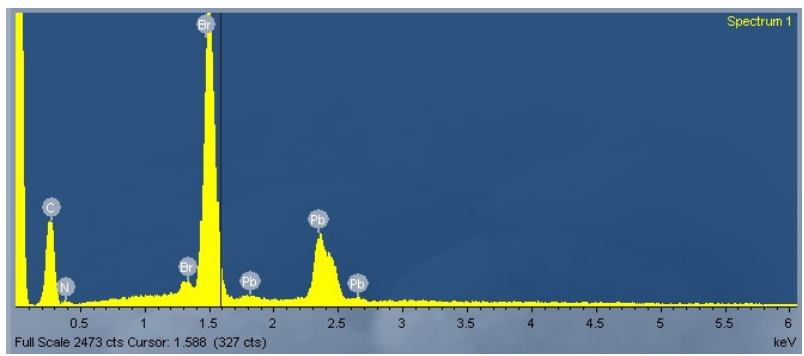

**EDX for mixture of quasi-2D HOIS: (Oleylamine-H)$_2$PbBr$_4$ and 3D (Methylamine-H)PbBr$_3$.**

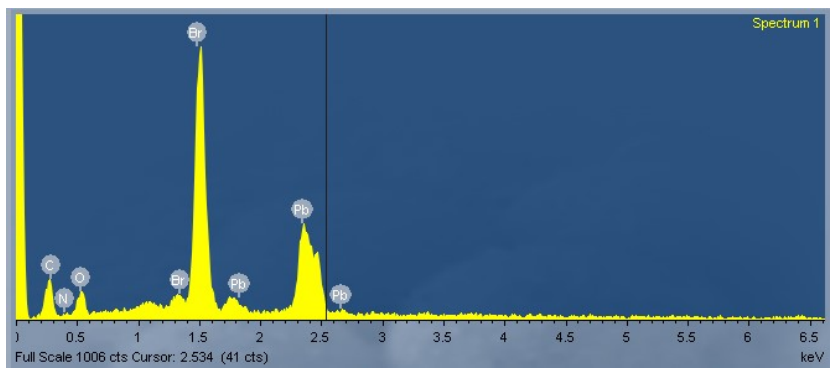

**Figure S1:** EDX images for various samples presented in this work: 3D, 2D based on various amines and quasi-2D/3D mixture based on oleylamine (2D) and methylamine(3D).

**FIGURE S2**

I-V characteristic curve of the sample DP53, which was a one deposited as one layer of DP53 on ITO and contacted with Cu -wire in the form of elastic spring. The potential and current axes have been negated values so that when electron flow from the Ga/In alloy into DP53, then the current is taken to be positive and voltage positive.

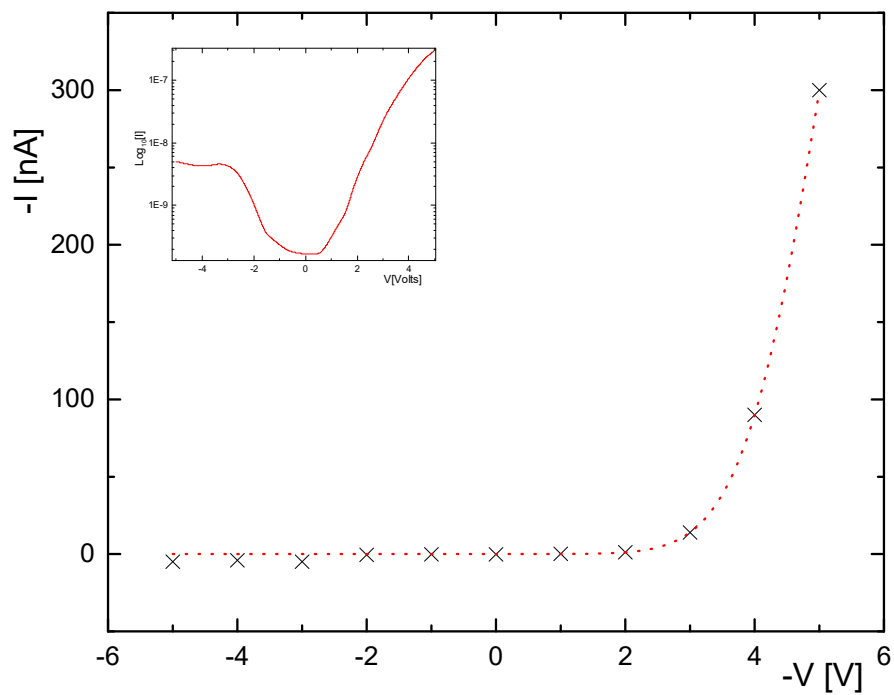

**Figure S2**. I-V characteristics of the diode ITO/DP53/Ga-In under low ambient laboratory light. Fitted line is for illustration purposes only. Inset shows the same for y as logarithmic axis.

**FIGURE S3**

SEM images in the range of 10μm-20μm showing the morphology of the 3D, quasi-2D and 2D HOIS..

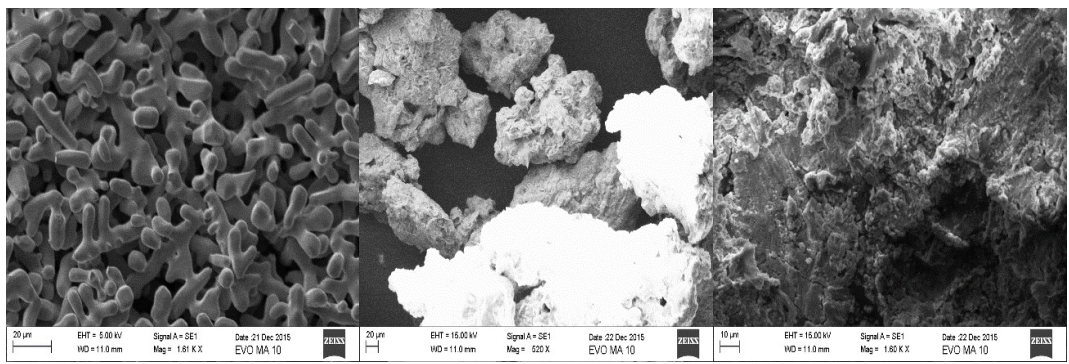

*Figure S3: SEM image of (left) 3D (Methylamine-H)PbBr$_3$, (center) 2D (Oleylamine-H)$_2$PbBr$_4$, (right) DP53 (mixture of the other two samples in a 3D:2D weight ratio 4.7:1).*

*Figure S4:*

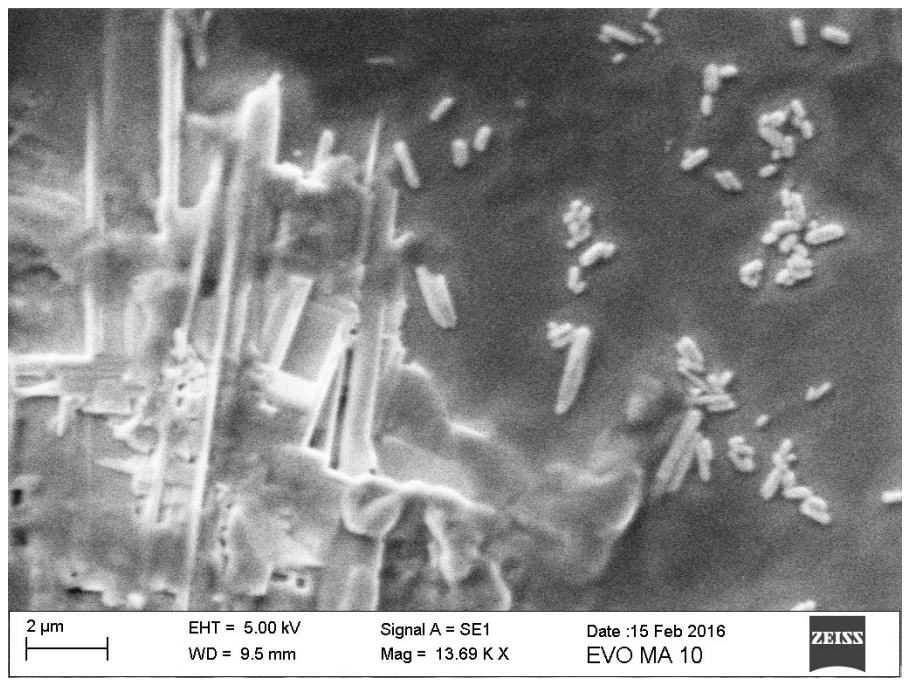

*Figure S4: SEM image of DP106 after heating at 115$^o$C for 30 minutes.*

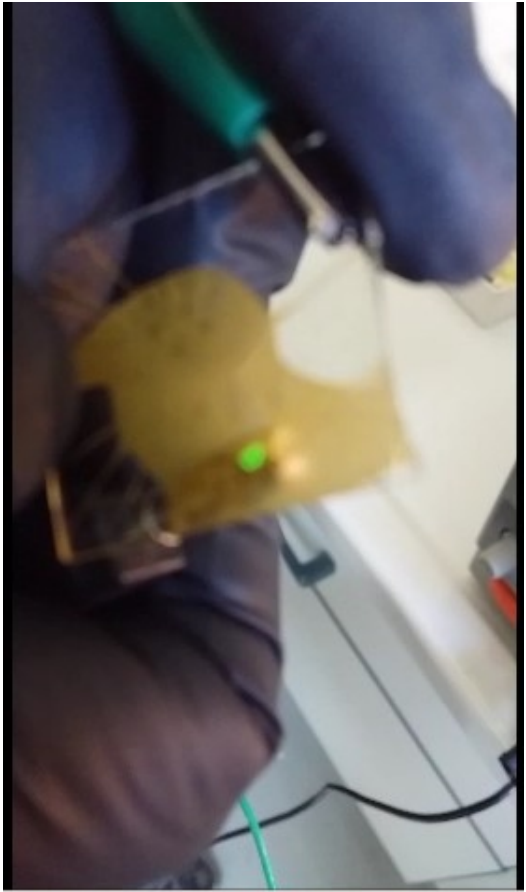

*DP53*

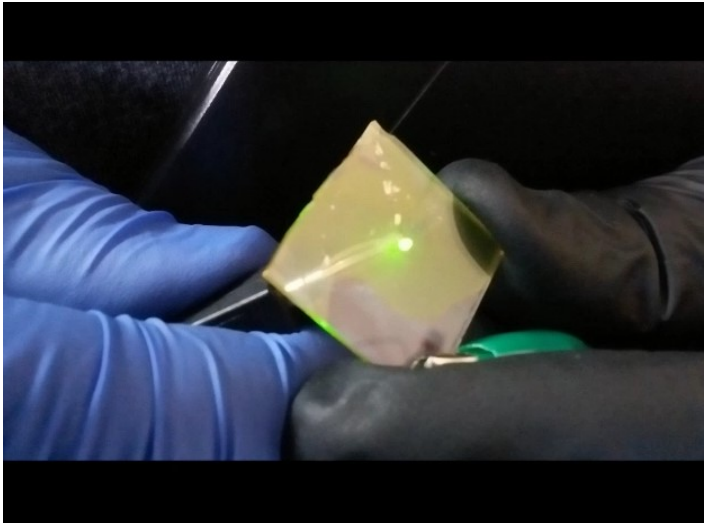

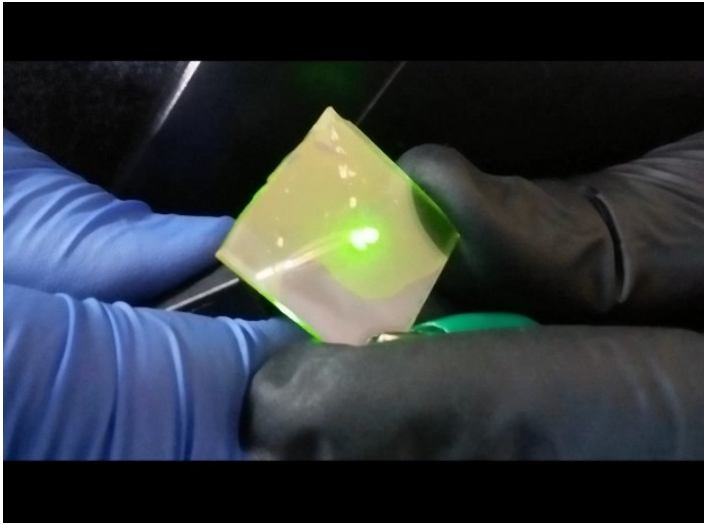

**DP80**

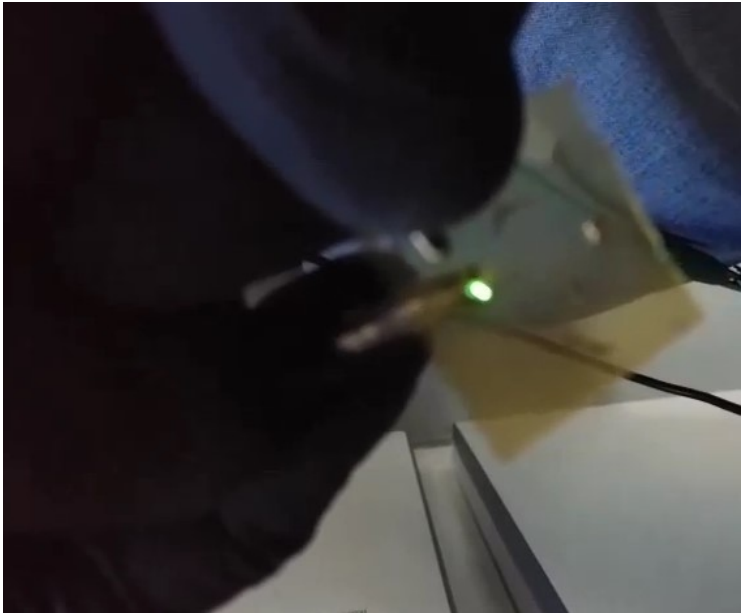

**DP106**

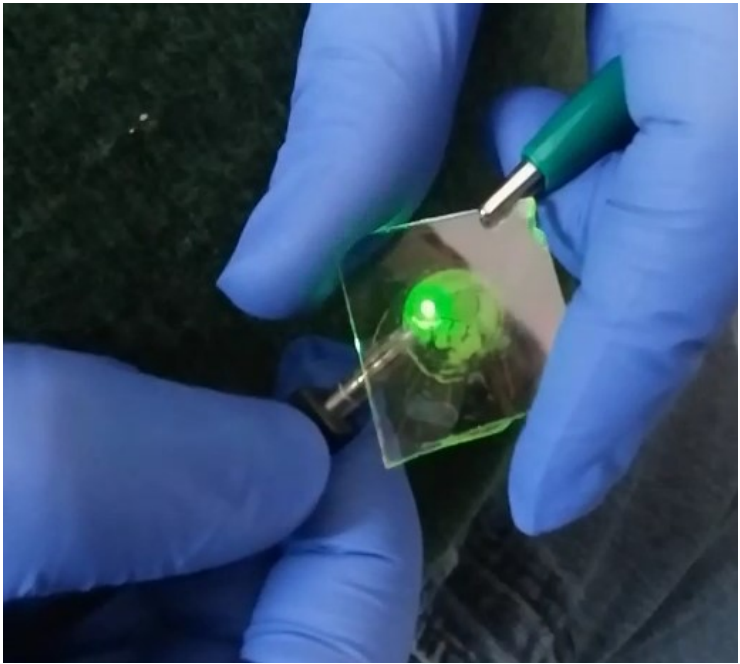
**DP112**